\documentclass{article}

\usepackage{PRIMEarxiv}

\usepackage[utf8]{inputenc} 
\usepackage[T1]{fontenc}    
\usepackage{hyperref}       
\usepackage{url}            
\usepackage{booktabs}       
\usepackage{amsfonts}       
\usepackage{nicefrac}       
\usepackage{microtype}      
\usepackage{lipsum}
\usepackage{fancyhdr}       
\usepackage{graphicx}       
\graphicspath{{media/}}     
\RequirePackage{algorithm}
\RequirePackage{algorithmic}

\title{Few-shot protein generation}

\author{
  Soumya Ram \\
  Massachusetts Institute of Technology \\
  Department of Biological Engineering\\
  Cambridge,MA,USA\\
   \And
  Tristan Bepler \\
  Simons Machine Learning Center \\
  New York Structural Biology Center \\
  New York, NY, USA\\
}

\begin{document}
\maketitle

\begin{abstract}
We present the MSA-to-protein transformer, a generative model of protein sequences conditioned on protein families represented by multiple sequence alignments (MSAs). Unlike existing approaches to learning generative models of protein families, the MSA-to-protein transformer conditions sequence generation directly on a learned encoding of the multiple sequence alignment, circumventing the need for fitting dedicated family models. By training on a large set of well curated multiple sequence alignments in Pfam, our MSA-to-protein transformer generalizes well to protein families not observed during training and outperforms conventional family modeling approaches, especially when MSAs are small. Our generative approach accurately models  epistasis and indels and allows for exact inference and efficient sampling unlike other approaches. We demonstrate the protein sequence modeling capabilities of our MSA-to-protein transformer and compare it with alternative sequence modeling approaches in comprehensive benchmark experiments.
\end{abstract}


\section{Introduction}
Protein engineering is the task of mutating proteins in order to achieve a desired function, and has numerous applications in medicine and sustainability \cite{jiri2021, Wilson2018-au}. However, designing these mutations is challenging, because it is difficult to infer their functional impact and the search space of possible sequence variants is combinatorialy large. A given mutation could cause a disproportionate effect due to being positioned within an active site or long-range interactions with other amino acids \cite{LiCata1995-po}. In addition, introducing multiple mutations simultaneously can have complex non-linear effects, called epistasis \cite{Starr2016-aa}. 

Supervised data on the functional impact of protein mutants is also limited. Acquiring supervised data means performing complicated and costly deep mutational scanning experiments. The number of variants that can be measured is limited by the assay throughput. For typical functional activity assays, this can restrict the number of feasibly measurable variants to hundreds or less. Generally, these experiments characterize the functional impact of point mutations, so experimental data on the impact of higher order mutants is even more rare \cite{Yang2019-tt, Riesselman2018-tr}.

In contrast, sequence data is plentiful for natural proteins. The number of known natural protein sequences has nearly tripled in the last 5 years, and continues to grow rapidly due to the falling cost of DNA sequencing. This proliferation has enabled new evolutionary sequenced modeling approaches for variant function prediction that circumvent the need for functional assay data. Classically, multiple sequence alignments (MSAs) of related protein sequences have proven useful for predicting protein contacts \cite{Morcos2011-gu, Balakrishnan2011-bp, Marks2012-sw, Seemayer2014-yh}, inferring  functional effects of mutants \cite{Levy2017-li}, and searching databases for related sequences \cite{Altschul1997-bi, Beckstette2009-ya}. Simple generative models learned from MSAs, such as position-specific scoring matrices (PSSMs) or profile Hidden Markov Models (pHMMs), are widely used protein sequence models \cite{Beckstette2009-ya}.

Recently, deep language models have achieved great success modeling proteins. Protein embedding models produce embeddings that surpass hand-crafted features in generalizing to downstream tasks \cite{bepler2018-le, Bepler2021-bu, Alley2019-wr, Elnaggar2021-nw, Rives2021-yw}.  Family-specific deep generative sequence models have also demonstrated impressive unsupervised variant function prediction on deep mutational scanning datasets \cite{Riesselman2018-tr, Shin2021-ra} and for predicting antibody escape mutations \cite{Hie2021-tq}. However, these models must be trained on a family-by-family basis. \cite{meier2021-la} recently observed that a transformer pre-trained as a masked language model on general MSAs is able to predict the effects of variants without retraining by feeding the family MSA directly as input to the model. However, this approach assumes an additive model of multiple variants and is unable to model insertions, limiting its usability. Inefficient inference in non-autoregressive models also creates challenges for their practical application to sequence generation.

Here, we introduce a new approach to learning generative models of proteins based on sequence-to-sequence learning. We newly formulate sequence modeling as a few-shot learning problem and train a single encoder-decoder model to accept a protein family represented as an MSAs as input and decode that MSA into a distribution over sequences from that family. We train the model on tens of thousands of multiple sequence alignments representing known protein families and evaluate it on unseen families held out from training. We achieve state-of-art performance on deep mutational scanning datasets and are able to model insertions and deletions in addition to substitutions. Our model also makes no additivity assumptions, allowing it to generalize well to highly diverse mutants. Unlike previous work, our model offers exact sampling and inference and crucially, does not require retraining for new families.

\section{Related Work}

Generative sequence models are widely used to represent protein families. PSSMs model each column in the MSA as independent distributions over amino acids. Profile HMMs model each amino acid as being generated conditioned on a hidden state corresponding to each column in the MSA, but this alignment is considered unobserved when calculating the probability of a new sequence. These simple models are widely used, because they need to be inferred from relatively small sets of sequences (often only 10s or 100s) and parameter inference needs to be performed for each set of proteins of interest. However, they are unable to model dependencies between amino acids.

Deep generative models have recently been explored for conditional and unconditional sequence generation, because they can model diverse protein sequences and capture high-order dependencies between amino acids. \cite{Madani2020} present an autoregressive transformer language model conditioned on taxonomic and functional labels. \cite{NEURIPS2019} develop a graph-based sequence generative model conditioned on protein tertiary structure. Other work has considered generative models for adaptively sampling from posterior distributions defined by function prediction methods, such as \cite{pmlr-v97-brookes19a}.  

Two of the proposed generative models for proteins require training a VAE on thousands of sequences for each new family \cite{Riesselman2018-tr, Mullaney2010-go}. The third generative model in the literature evaluated on variant function prediction, MSA Transformer, also can do few-shot prediction \cite{meier2021-la}. Compared to MSA Transformer, our formulation utilizes a decoder that attends over the multiple sequence alignment to explicitly model sequence probabilities autoregressively. In contrast, the MSA Transformer is a masked language model which cannot give a closed-form sequence probability, is unable to generalize to sequences with a different number of positions from the MSA, and does not permit efficient sampling.  

The addition of a decoder allows for many benefits in inference and sampling. Because of the decoder, our model is able to produce exact likelihoods for higher order mutants. Masked zero-shot language models such as MSATransformer and ESM-1v evaluate likelihoods for higher order mutants by assuming additivity \cite{meier2021-la}. However, these proteins tend to have strong epistatic effects so such an assumption is not ideal \cite{Starr2016-aa}. 
 
 MSA2Prot is also able to generate sequences with insertions as well, allowing it to model the entire protein landscape. This is especially valuable as indels comprise of 15-21 percent of human polymorphisms, and inferring their functional impact is important for understanding genes related to disease \cite{Mullaney2010-go}. 

Sampling-wise, our model produces exact samples while zero-shot masked language models \cite{meier2021-la} require Gibbs sampling. Gibbs sampling struggles with transitioning between high probability states that are connected by low probability paths. This is the case with many higher order mutants. For example, double mutants can be beneficial when each of the single mutants are not because of epistatic interactions. In addition, this issue is compounded by the fact that directed evolution experiments can require a large number (possibly thousands) of initial protein samples. Producing this many independent samples through Gibbs creates a large computational burden. 

Thus, MSA2Prot's formulation models epistatic effects from higher order mutants and indels, while also allowing for efficient and exact sampling and evaluation. 

Lastly, MSA2Prot can be used for adaptive sampling in settings with scarce protein fitness measurements as it generalizes from the input MSA without needing retraining. This is in contrast to methods such as \cite{Brookes-1} which continually require the generator to be retrained. Thus, MSA2Prot is able to combine the data efficiency of simple models such as HMMs with the expressive generative capability of deep learning models. 

\section{Methods}

\begin{figure}[h!]
\centering
\label{main_figure}
  \caption{MSA2Prot Architecture}
  \includegraphics[width=0.60\textwidth]{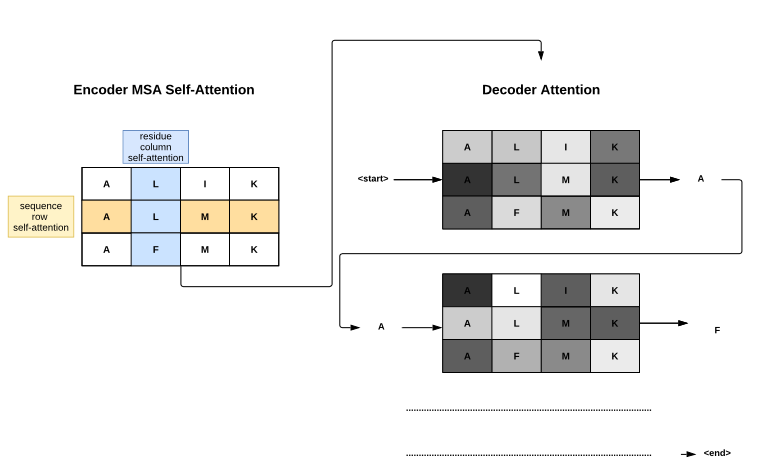}
\end{figure}

Let $x_{i,j}$ , $i \in 1...N$ , $j \in 1...M$, denote the amino acid or gap token at the $j$’th position of the $i$’th sequence in a multiple sequence alignment (the query MSA) with N sequences and M columns. Let $y_i$ , $i \in 1...L$ , be the token for the $i$’th amino acid in the target protein sequence. Our goal is to learn the probability distribution over target protein sequences ($y$’s) from protein families represented by their members in a multiple sequence alignment ($X$’s), $p(y | X)$, given a training set of MSA, target protein pairs ($X^k$,$y^k$) where $k \in 1...K$ denotes the $k$’th pair. Here, target protein $y^k$ is a member of the same family as the sequences in $X^k$ that is not itself in $X^k$.
In order to generalize across input MSAs, we propose an MSA-encoder, sequence-decoder neural network architecture. The protein MSA is encoded using an MSA transformer with axial self-attention on the rows and columns, into context-aware vector representations, $z_{i,j}$, for each token in the MSA, and then the target sequence is decoded by attending to the learned MSA representations in a decoder transformer with cross attention to the encoded MSA.This is illustrated in Figure \ref{main_figure}

\subsection{The Transformer MSA Encoder}

The MSA encoder accepts a multiple sequence alignment as input and returns a vector representation for each position in the multiple sequence alignment, $ z_{i,j} \in R^d $, where $d$ is the dimension of the learned embedding. This encoder is parameterized as a stack of transformer layers with axial attention over the rows and columns of the MSA. Before being processed by the transformer stack, the input tokens are embedded into vectors in $R^d$ and augmented with a random Fourier projection of the column index as a positional embedding, also in $R^d$. No positional embedding is used for the rows of the MSA (the sequence index), because the ordering of sequences in an MSA is arbitrary and we wish for the MSA encoder to be invariant to the specific ordering of sequences in the input.

\paragraph{The input embedding} is formed by adding a learned embedding for each amino acid to the random Fourier feature embedding of the column index as follows. First, the amino acid token is embedded
$$x_{i,j}^{aa}=W_{x_{i,j}}, x_{i,j}^{aa} \in R^d ,W \in R^{K \times d}$$              
where $W$ is a matrix of learnable amino acid embeddings of dimension $d$, K is the size of the vocabulary (22 in the case of 20 amino acids plus gap and start/end tokens), and $x_{i,j}$ indicates the amino acid at position $i,j$ of the MSA. Next, the column index is embedded
$$x_{i,j}^{pos}=W^{pos}\cos(rj+b),$$
$x_{i,j}^{pos} \in R^d, r \in R^d, b \in R^d, W^{pos} \in R^{d \times d} $, where $W^{pos}$ is a learnable matrix, $r$ is a random vector drawn from Normal(0, 1), and $b$ Is a random vector drawn from Uniform(0, 2$\pi$). The input embedding is then formed by
$$ z_{i,j}=x_{i,j}^{aa} + x_{i,j}^{pos} $$

\paragraph{The MSA encoder transformer layers} are composed of axial self-attention along the rows and columns of the MSA matrix followed by a fully connected layer. Each operation is preceded by layer normalization and uses residual connections (Algorithm~\ref{alg:msa_encoder}).

\begin{algorithm}[tb]
   \caption{Transformer MSA Encoder with Axial Attention}
   \label{alg:msa_encoder}
\begin{algorithmic}
   \STATE {\bfseries Input:} MSA $z_{i,j}  \in R^d,  i \in 1...N, j \in 1...M$
   
   \STATE {First, attention within each row}
   \STATE {$z_{i,j}' = LayerNorm(z_{i,j})$}
   \STATE {$z_{i,j}' = RowAttention(z_{i,j}')$}
   \STATE {$z_{i,j} = z_{i,j} + z_{i,j}'$}
   
   \STATE {Next, attention within each column}
   \STATE {$z_{i,j}' = LayerNorm(z_{i,j})$}
   \STATE {$z_{i,j}' = ColumnAttention(z_{i,j}')$}
   \STATE {$z_{i,j} = z_{i,j} + z_{i,j}'$}
   
   \STATE {Finally, the feed forward layer}
   \STATE {$z_{i,j}' = LayerNorm(z_{i,j})$}
   \STATE {$h_{i,j} = GeLU(Linear(z_{i,j}' )) , h_{i,j} \in R^{4d}$}
   \STATE {$z_{i,j}' = Linear(h_{i,j})$}
   \STATE {$z_{i,j} = z_{i,j} + z_{i,j}'$}
   
   \STATE {\bfseries Return:} $z_{i,j} \in R^d$
   
\end{algorithmic}
\end{algorithm}

Within row and within column attention are computed efficiently by calculating the multi-headed attention operation over the rows or columns batch-wise. That is, given a batch of intermediate MSA representations, Z, with dimensions BxNxMxd, where B is the number of MSAs in the batch, we calculate per-row self-attention by treating rows as part of the batch dimension, BNxMxd, and we calculate per-column self-attention by treating columns as part of the batch dimension, BMxNxd.

\subsection{The Transformer MSA Decoder}

\paragraph{The sequence decoder transformer layers} are composed of causal self-attention, cross-attention to the MSA representations, and fully connected layers with layer normalization and residual connection for each block (Algorithm~\ref{alg:msa_decoder}). As is typical in transformer decoders, the causal self-attention mask ensures that the output representation for each position of the sequence in the decoder is only a function of previous positions in the sequence and the MSA representations. Before being passed into the decoder transformer layers, the target sequence is embedded following the same scheme as the input embedding for the MSA along the column dimension, $a_k$. When decoding, the target sequence is padded to begin with a start token and end with a stop token.

\begin{algorithm}[tb]
   \caption{Transformer Sequence Decoder with MSA Cross Attention}
   \label{alg:msa_decoder}
\begin{algorithmic}
   \STATE {\bfseries Input:} MSA representations $z_{i,j} \in R^d, i \in 1...N, j \in 1...M$, and target sequence $a_k \in R^d, k \in 1...L$
   
   \STATE {First, casual self-attention within the target sequence}
   \STATE {$a'_k=LayerNorm(a_k)$}
   \STATE {$a'_k=CausalSelfAttention(a'_k)$}
   \STATE {$a_k=a_k+a'_k$}
   
   \STATE {Next, cross attention against the MSA representations}
   \STATE {$z_{i,j}'= LayerNorm(z_{i,j})$}
   \STATE {$a'_k=LayerNorm(a_k)$}
   \STATE {$a'_k=MSACrossAttention(a'_k, z_{i,j}')$}
   
   \STATE {Finally, the feed forward layer}
   \STATE {$a'_k = LayerNorm(a_k)$}
   \STATE {$h_k = GeLU(Linear(a'_k )), h_k \in R^{4d}$}
   \STATE {$a'_k = Linear(h_k)$}
   \STATE {$a_k=a_k+a'_k$}
   
   \STATE {\bfseries Return:} $a_k \in R^d$
   
\end{algorithmic}
\end{algorithm}

In the MSA cross attention module, each position of the target sequence attends over the complete MSA representations for each attention head (i.e., LxNxMxH, where H is the number of heads). We compute this efficiently by flattening the MSA representations along the row and column dimensions such that each MSA representation, $z_{i,j}$, is a single key in the cross attention module.

Decoding the target sequence is performed using the decoder representations by learning a transformation from $a_k$ to the probability distribution over the $k+1$th token. We formulate this as a linear transformation of $a_k$ into a vector of dimension equal to the number of tokens followed by softmax to give the probability of each token, $p(y_{k+1} | a_k) = p(y_{k+1} |  y_{1...k}  , X)$.

\subsection{Loss, hyperparameters, and training}

\paragraph{Loss function.} We fit the parameters of the model to minimize the negative log-likelihood of the target sequences conditioned on their family MSAs. Given $K$ MSA, target sequence pairs, this loss is 
$$ \mathcal{L} = - \sum_{i=1}^{K} \sum_{k=1}^{L^i} \log p(y_k^i | y_{1...k-1}^i, X^i) $$

\paragraph{Model hyperparameters.} Here, we train models with the following specific hyperparameters. We use 6 encoder and decoder layers each with hidden dimension ($d$) of 768. The MSA encoder uses 12 attention heads and the decoder uses 8 attention heads.

\paragraph{Training details.} We train the model using the ADAM variant of stochastic gradient descent using a linear ramp up, square root decay learning rate scheduler using a learning rate of 0.0001 with 4,000 warmup steps. We use a total minibatch size of 256 spread over 16 GPUs using distributed training. Each GPU process minibatches of size 1 with gradient accumulation over 16 steps to give the total effective minibatch size.  In order to reduce GPU RAM consumption, sequences and MSAs are randomly sampled to a maximum length of 402 tokens during training. Furthermore, MSAs are randomly downsampled to contain between 1 and 50 sequences. During training, we monitor the loss on a validation set of MSAs and stop training when the validation loss stops decreasing.

\paragraph{Datasets.} We train and evaluate the model on full Pfam family alignments. In order to evaluate the performance of the model on unseen families, we split the Pfam sequences at the family level into 10,593 training, 563 validation, and 2,654 test families.

\section{Results}

\subsection{Few shot protein generation of unseen protein families}

We first examine the ability of our model to generate new proteins by generalizing from few protein sequences representing a protein family to unseen members of the family. We calculate the perplexity of sequences in the validation set families given an increasing number of randomly selected observed members, not including the target sequence. This allows us to understand the ability of our model to extrapolate evolutionary landscapes from a small number of observations. We compare our model with two widely used protein statistical sequence models: position-specific scoring matrices (PSSMs) and profile HMMs (pHMMs).

\begin{table}[t]
\centering
\caption{Perplexities of generative models on the Pfam validation families as the number of observed sequences increases. For each family, a target sequence is sampled and then an MSA with the specified number of sequences is sampled from the remaining sequences in the family. A PSSM or pHMM is then fit to the MSA and used to calculate the perplexity of the target sequence or for our model. For our model, the feed the MSA into the trained encoder and calculate the perplexity of the target sequence using the trained decoder. We repeat this 1,000 times for each family and report the average perplexity over trials and families. Our model achieves lower perplexity at every MSA size and even outperforms the profile HMM with 10x as many sequences.}
\label{tab:generative_perplexity}
\vskip 0.15in
\begin{center}
\begin{small}
\begin{sc}
\begin{tabular}{lrrrrr}
\toprule
& \multicolumn{5}{c}{Number of sequences in MSA} \\
Method & 1 & 10 & 25 & 50 & 100 \\
\midrule
PSSM & 14.30 & 8.43 & 7.17 & 6.64 & 6.20 \\
pHMM & 12.30 & 6.65 & 5.85 & 5.51 & 5.32 \\
Ours & \textbf{9.48} & \textbf{4.44} & \textbf{3.62} & \textbf{3.20} & \textbf{2.92} \\
\bottomrule
\end{tabular}
\end{sc}
\end{small}
\end{center}
\vskip -0.1in
\end{table}

As expected, we observe that all models’ ability to generalize to unseen family members improves as the number of observed family members increases (Table~\ref{tab:generative_perplexity}). Furthermore, our model dramatically outperforms PSSMs and profile HMMs across all MSA sizes. Remarkably, our model scales much better with additional data, even outperforming PSSMs and pHMMs with 10x fewer sequences. This demonstrates that our model learns how to extrapolate from small number of sequences by better capturing evolutionary priors over sequences .

A major challenge in protein engineering is navigating the enormous search space of possible sequence variants, because the space of sequence variants increases exponentially with the number of sites . For example, if we consider all 20 amino acids at 10 positions, the number of unique sequences is 2010 which is greater than ten trillion. At 65 sites, the space of possible sequences exceeds the number of atoms in the universe. However, the vast majority of these variants are not functional ($<$1\% in typical mutagenesis experiments). Therefore, homing in on only the space of viable protein variants is critical for efficiently and feasibly searching sequence space. Perplexity represents the number of amino acids that would need to be guessed from uniformly to find the correct amino acid, therefore, it is the size of the reduced alphabet learned by the model. On this basis, our model can produce an enormous reduction in library size for protein engineering. Using the 10 sites example, the pHMM perplexity of 5.3 yields a library size of 18 million which our model reduces to only 42,000, more than an order of magnitude reduction in library size over the pHMM and about 8 orders of magnitude better than random search. This improvement is even more extreme when considering more sites for mutation.

Compared to the masked language models described, MSA2Prot offers exact sampling through the use of a decoder. This sidesteps computationally intensive Gibbs sampling. 

\subsection{MSA2Prot Log-Likelihood Predicts Variant Function}
\subsubsection{Point Mutants}
We evaluate our model on the protein mutation datasets developed by \cite{Riesselman2018-tr}. These datasets consist of mostly single mutant deep mutational scans, and a few double mutant deep mutational scans. 

As a baseline, we compare to an unconditional language model trained on the same dataset and MSA2Prot predictions with only the wild-type sequence. We also compare to to state-of-art methods for protein fitness prediction. These include PSSMs, pHMMs, EVmutation, DeepSequence, ESM-1v, and MSATransformer. 

PSSMs estimate the probability of each amino acid at a given position and assume independence between different positions. EVmutation is a Potts model that estimates the pairwise interaction terms between residues. pHMMs model amino acids as being generated conditioned on a hidden state of the respective MSA column. DeepSequence trains a generative model on a large multiple sequence alignment for a given protein. ESM-1v and MSATransformer are masked language models, where ESM-1v models protein sequences whereas MSATransformer models multiple sequence alignments. 

\begin{table*}[t]
\centering
\caption{Comparison of MSA2Prot to unsupervised protein fitness predictors. New baselines include an unconditional language model and MSA2Prot with only the wild-type sequence. Spearman Rho is averaged across 40 deep mutational scanning datasets. HIS7 YEAST is omitted due to its large size.}
\label{tab:dms}
\vskip 0.15in
\begin{small}
\begin{sc}
\begin{tabular}{lr}
\toprule
Method & DMS Spearman Rho  \\
\midrule
MSA2Prot & 0.536  \\
MSATransformer \cite{meier2021-la} & \textbf{0.548} \\
ESM-1v (finetuned) \cite{Meier2021.07.09.450648} & 0.538 \\
DeepSequence \cite{Riesselman2018-tr} & 0.513 \\
ESM-1v (zero-shot) \cite{Meier2021.07.09.450648} & 0.508 \\
EVMutation \cite{Hopf2017-ia} & 0.506  \\
PSSM & 0.460 \\
MSA2Prot wildtype & 0.272  \\
Unconditional LM & 0.091  \\
\bottomrule
\end{tabular}
\end{sc}
\end{small}

\vskip -0.1in
\end{table*}

\begin{figure}[h!]
\centering
\label{dms_figure}
  \caption{Spearman Rho Across 40 DMS Datasets}
  \includegraphics[width=0.42\textwidth]{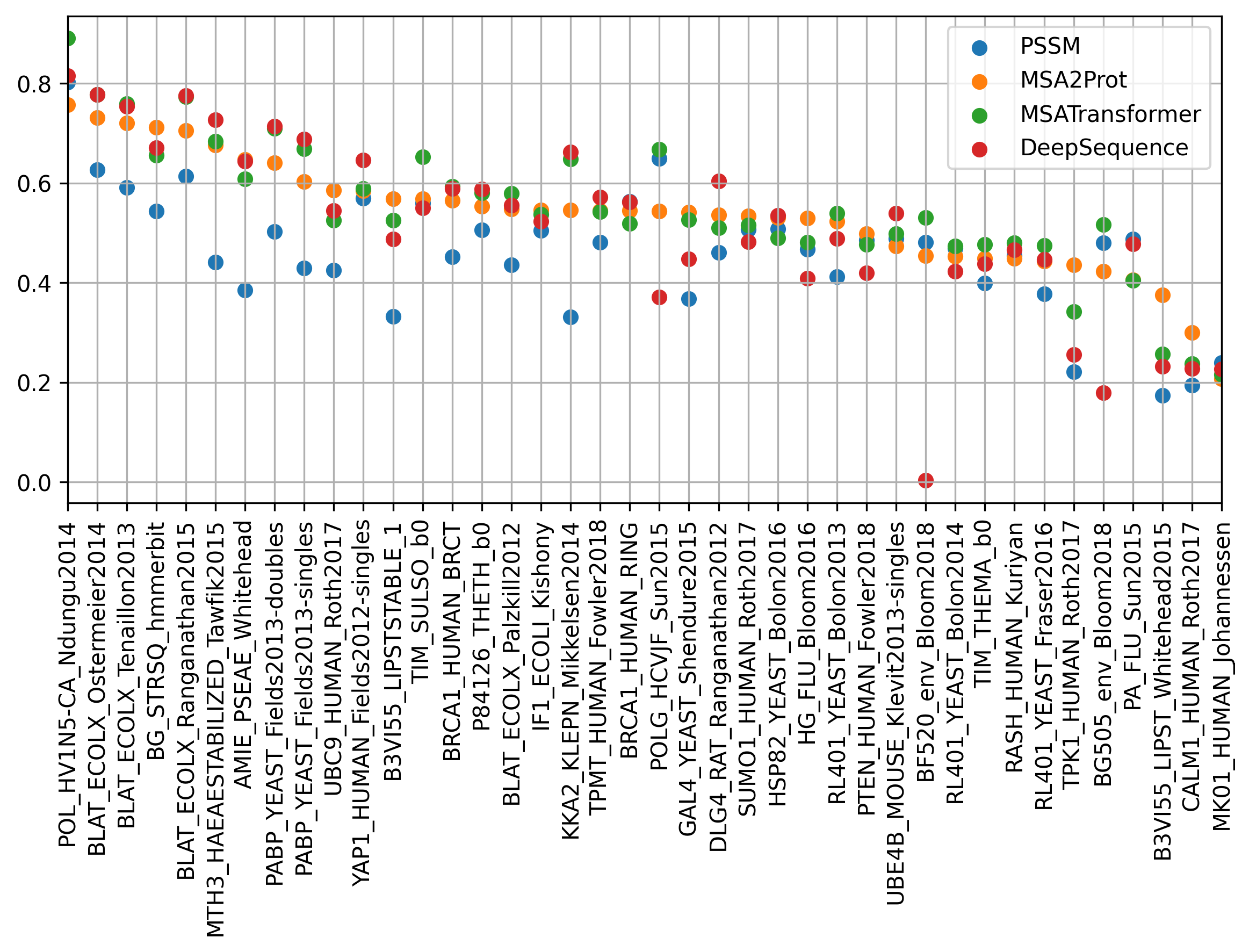}
\end{figure}

Compared to the above methods, MSA2Prot displays state-of-the-art performance. Comparisons with the unconditional language model as well as the wild-type only MSA2Prot predictions indicate that the inclusion of the MSA is a key driver of this performance. The averages across the datasets are shown in Table \ref{tab:dms}, and a dataset-specific breakdown is shown in Figure \ref{dms_figure} . 

\subsubsection{Higher Order Mutants}
Given that 37 of above 40 DMS datasets consist of single mutants, we evaluate MSA2Prot on a dataset of chorismate mutase sequences \cite{Russ2020-tq}. This dataset is highly diverse, with sequence divergence in the test set ranging from 14 percent to 82 percent. 

\begin{table}[t]
\centering
\caption{Baseline spearman rho on chorismate mutate across methods. HMM, PSSM, and Potts were fit to the MSA of natural sequences created by \cite{Russ2020-tq}. For MSA2Prot, sequences ensembled with MSA size 100 and were reweighted with sequences above 0.7 percent similarity considered to be neighbors. For MSA Transformer, the best value across sequence reweighting with different thresholds,random sampling, and the raw msa is reported.}
\label{tab:chorismate_baseline}
\vskip 0.15in
\begin{small}
\begin{sc}
\begin{tabular}{lr}
\toprule
Method & Spearman Rho  \\
\midrule
MSA2Prot & \textbf{0.41}\\
MSATransformer & 0.12  \\
HMM & 0.38  \\
Potts & 0.41  \\
PSSM & 0.39 \\
\bottomrule
\end{tabular}
\end{sc}
\end{small}

\vskip -0.1in
\end{table}

We find that MSA2Prot outperforms MSATransformer substantially, as shown in Table \ref{tab:chorismate_baseline}. This is likely because MSATransformer uses additive scoring to assess the likelihood of higher order mutants, whereas MSA2Prot’s decoder is able to exactly model epistasis.

\begin{table}[t]
\centering
\caption{Spearman Rho for standard MSA, MSA filtered for variants above fitness level 0.4, and MSA filtered for variants above fitness level 0.7. All MSAs were reweighted with sequences above 0.7 percent similarity considered to be neighbors. For the ensemble, MSAs of size 100 were sampled. MSA2Prot is able to utilize negative samples, the 350 worst performing training sequences, even when they have no predictive value on their own. MSA2Prot is also able to generalize from the distribution of high-performing mutants.}
\label{tab:chorismate_trials_spearmanrho}
\vskip 0.15in
\begin{small}
\begin{sc}
\begin{tabular}{lrrrrr}
\toprule
& \multicolumn{5}{c}{Fitness Level of MSA} \\
Method & None & 0.4 & 0.7 \\
\midrule
Ensemble & 0.41 & 0.57 & 0.58 \\
No Ensemble & 0.37 & 0.58 & 0.60 \\
Ensemble w/ Neg Samples & 0.55 & 0.58 & 0.59 \\
No Ensemble w/ Neg Samples & 0.50 & 0.59 & 0.58 \\
PSSM & 0.39 & 0.50 & 0.50 \\
HMM & 0.38 & 0.48 & 0.48 \\
Negative Samples Only & 0.00 & 0.00 & 0.00\\
\bottomrule
\end{tabular}
\end{sc}
\end{small}

\vskip -0.1in
\end{table}

In addition, we find that MSA2Prot is able to generalize from a distribution of high-performing mutants, as shown in Table \ref{tab:chorismate_trials_spearmanrho}. Spearman rho on the test set dramatically improves after the training MSA is filtered to include only a few hundred high-performing sequences. Lastly, we find that MSA2Prot is able to harness low-performing variants to significantly improve accuracy. Although these low-performing variants have no predictive power on their own, we use them as negative samples by subtracting the likelihood of a sequence from the low-performing MSA from the likelihood of a sequence from a standard MSA.

Often, combing the literature for a given protein will yield a list of high and low performing mutants. However, given that experimental setups differ, there may not be consistent fitness measurements. MSA2Prot is an ideal candidate for this situation, given its ability to harness both high and low performing variants without explicit functional measurements. 

MSA2Prot also offers exact generation conditioned on multiple attributes. Given a protein sequence, the probability distribution over the next residue can be obtained by adding and re-normalizing the marginals of two MSAs, each representing different attributes.

\subsubsection{Indels}
\begin{table}[t]
\centering
\caption{Spearman Rho across methods on beta-lactamase indel dataset}
\label{tab:blat_ecolx}
\vskip 0.15in
\begin{small}
\begin{sc}
\begin{tabular}{lr}
\toprule
Method & Spearman Rho  \\
\midrule
MSA2Prot & 0.46 \\
DeepSequence & 0.45  \\
HMM & 0.52  \\
\bottomrule
\end{tabular}
\end{sc}
\end{small}
\vskip -0.1in
\end{table}

We evaluate MSA2Prot on \cite{Gonzalez2019-ga}'s dataset of 262 deletions and 4422 insertions. We benchmark against \cite{Riesselman2018-tr} and a HMM. Table \ref{tab:blat_ecolx} indicates that MSA2Prot achieves comparable performance with \cite{Riesselman2018-tr}, even though it does not require retraining.

\subsection{Adaptive sampling with MSA2Prot}

\begin{table}[t]
\centering
\caption{Median log likelihood for test proteins with fitness greater than 0.4. Log likelihoods of test proteins were evaluated from an MSA with no filtering, an MSA containing training sequences with fitness greater than 0.4, and an MSA containing training sequences with fitness greater than 0.7. All MSAs were reweighted with sequences above 0.7 percent similarity considered to be neighbors. An input MSA with higher fitness leads to a higher likelihood on high-performing test set variants}
\label{tab:chorismate_perplexity}
\vskip 0.15in
\begin{small}
\begin{sc}
\begin{tabular}{lr}
\toprule
MSA Fitness & Log Likelihood  \\
\midrule
None & -124.36 \\
0.4 & -112.52  \\
0.7 & -108.92  \\
\bottomrule
\end{tabular}
\end{sc}
\end{small}

\vskip -0.1in
\end{table}

In addition to offering high predictive accuracy through its log-likelihood, MSA2Prot is able to generatively extrapolate from high-performing mutants. We find that the high-performing sequences in the test set become several orders of magnitude more likely as the training msa is filtered to include higher performing variants. This result is shown by Table \ref{tab:chorismate_perplexity}.

\begin{figure}[h!]
\centering
\label{chain1_figure}
  \caption{Change in Fitness on Chorismate Mutase Dataset Through Adaptive Sampling. Sampled sequences replace MSA if they have a higher predicted fitness, which is approximated by a Random Forest Regressor.Probability distributions across each group of 500 sequences are shown}
  \includegraphics[width=0.42\textwidth]{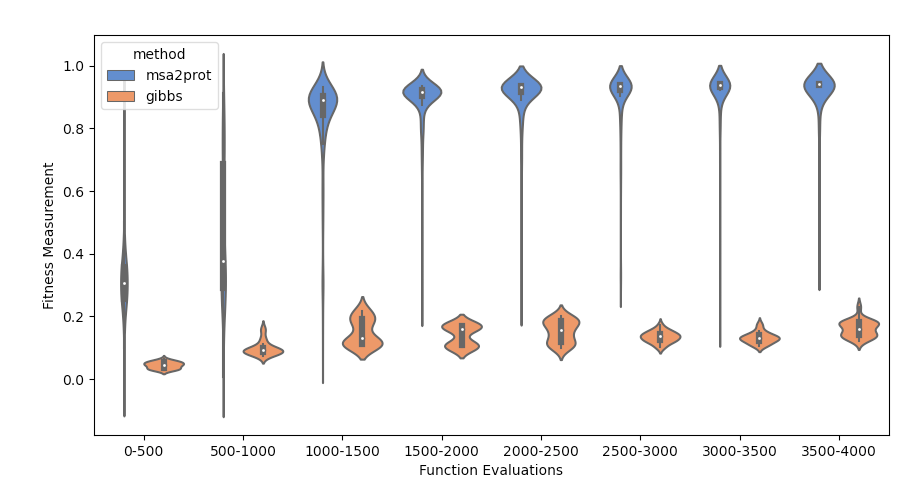}
\end{figure}

We also test MSA2Prot’s ability to adaptively sample high fitness variants given black-box oracle,approximated by a Random Forest Regressor. We randomly sample 100 sequences from the training MSA to form an initial MSA. We sample sequences, and update the MSA if the regressor predicts the sampled sequence has a higher fitness than the minimum fitness sequence in the MSA.  The results, shown in Figure \ref{chain1_figure}, indicate that MSA2Prot is able to effectively generate strong mutants. MSA2Prot initially generated sequences with a fitness of 0.3, which is the 56th percentile of the training distribution. After several thousand updates, MSA2Prot generated sequences with a fitness of 0.9, which is in the 89th percentile of the training distribution.For comparison, Gibbs sampling initially generated a sequence with a fitness of 0.03, which is in the 39th percentile of the training distribution. After the same number of function evaluations, Gibbs sampling generated a sequence with fitness 0.3, which is in the 56th percentile of the training distribution. Thus, MSA2Prot displayed considerably stronger performance. Compared to standard adaptive sampling methods, MSA2Prot offers the benefit of not requiring training. Instead, MSA2Prot relies on its few-shot generalization abilities. Thus, MSA2Prot offers a computationally efficient alternative for protein sequence design.


\section{Conclusion}

We present MSA2Prot, a transformer-based model for protein sequence generation directly conditioned on related protein sequences represented as a multiple sequence alignment. This model allows for efficient sampling and exact log-likelihood evaluation for sequences including insertions and deletions as well as substitutions without any need for training or fine-tuning on an MSA-by-MSA basis. The MSA2Prot model generalizes well to unseen families and generates to unseen family members with higher probability than other family generative models, especially when MSAs are small. Furthermore, we demonstrate that this model achieves state-of-the-art performance in variant function prediction on diverse mutagenesis datasets. Finally, MSA2Prot unlocks new capabilities in adaptive sequence generation by allowing the sampler to be focused only on functional variants without expensive retraining.



\bibliographystyle{unsrt}  
\bibliography{references}

 \newpage
 \appendix
 \onecolumn

 \section{DMS experiment}
\cite{Riesselman2018-tr} compiled a set of 41 deep mutational scanning experiments, almost all single-point mutations, and we used the MSAs provided by them. MSA filtering is described section C below. HIS7 YEAST was omitted from the comparison due to its large size of 496,000. 

 \section{Chorismate Mutase Dataset}
\cite{Russ2020-tq} created a MSA of 1130 natural chorismate mutase enzymes. They sampled 1618 sequences from Potts models trained on this MSA. The MSA of natural enzymes was used as a training MSA, and the 1618 variants were used as a test set. Fitness was measured for sequences in the training and test set. 
 
 \subsection{Spearman Rho Comparison Across Methods}
 For MSA2Prot,the MSA was ensembled by sampling groups of 100 sequences, where a sequence's probability of selection was inversely proportional to its number of neighbors. Two sequences were considered neighbors if their sequence similarity was greater than 0.7.The total number of groups of 100 was determined by the number of effective sequences in the MSA. For MSATransformer, the best spearman rho across sequence reweighting with different thresholds,random sampling, and the raw msa is reported. Perplexity was calculated using masked marginals and the wild-type sequence was chosen to be E.coli. 
 All other baselines were trained on the natural enzyme MSA without any filtering. ESM-1v was omitted as a baseline due to the many insertions in the MSA, relative to the wild-type.
 
 \subsection{Spearman Rho Comparison Across MSA Fitness}
 The training MSA was filtered to only include sequences above a threshold. HMMs and PSSMs were trained on this MSA without any further filtering. If ensembled, MSAs of size 100 were sampled from the training MSA, where a sequence's probability of selection was inversely proportional to its number of neighbors. Two sequences were considered neighbors if their sequence similarity was greater than 0.7. The total number of sequences sampled was determined by the number of effective sequences in the MSA. For negative sampling, an MSA was creating with the 350 worst-performing training sequences. To generate spearman rho values, the predicted likelihood of a sequence was the likelihood under the negative MSA subtracted from twice the likelihood of the sequence under the standard MSA. 
 
 \subsection{Log Likelihood Comparison Across MSA Fitness}
 Three MSAs were created, an MSA with no filtering, an MSA containing training sequences with fitness greater than 0.4, and an MSA containing training sequences with fitness greater than 0.7. Each of these MSAs were further reweighted as described above, and the log likelihood of test proteins with fitness greater than 0.4 was measured. 

 \subsection{Adaptive Sampling}
 A random forest regression model with 1000 estimators was trained on the training and test MSAs as a baseline. 
 100 sequences from the training MSA were randomly sampled to form an MSA. Sequences from this MSA were sampled and evaluated using the regressor. If a sampled sequence had a higher fitness than the minimum fitness of the sequences in the MSA, then the minimum fitness sequence in the MSA would be replaced by the sampled sequence.
 
 The last 3 sequences from the adaptive sampling experiment are shown below.

 \scalebox{0.7}{LALGQLADERRDRDAQALALIDARQALAERVREVKRAAGAEVTDATRERAVLDRAEAGAVPQGEQNGRQGDSVLPELFAALYAQGRQAEALRARW}
 \scalebox{0.7}{LHERLAIQVRMALDRELLFLIDARLTFAQQIGQIKRGKGAPVSDPSRQRVLQERAEANARAPQLAQPQLLVDLFERLFARSHTAERIQKRL}
 \scalebox{0.7}{LELSYARQQQIDVDRRLLAMLAQRLPLAQQINNMKRSGGALVADPLRSRVIEERSNLMAATHSSDEMPFISDQFGRLFERVRNAERQARRR}

 \section{Beta-lactamase dataset}
 This deep mutational scanning dataset includes 4422 single insertions and 262 single deletions. An MSA was created using MMSeqs2, and was filtered and ensembled as described above. The HMM was fit on the MSA, without any filtering.

 \section{Details of the MSA sampling strategies experiments}
 
 We experimented with several strategies for MSA selection. We varied MSAs by size, ensembled MSAs, and combined MSAs with different sequence similarities.We tested these strategies on a subset of five proteins from the DeepSequence dataset. We report the average spearman rho across all of the measurements for each protein.

We found that increasing MSA size, averaging predictions across multiple sampled MSAs, and averaging predictions from MSAs sampled with different sequence similarities all lead to improved performance. Thus, our final MSA strategy consisted of averaging six MSAs of size 100. For each sequence similarity filter, 1.0, 0.7, and 0.4, we sample a MSA twice.

For a given sequence similarity threshold, we removed sequences where the percentage of amino acids that differ from the wild type protein was greater than the threshold. A probability distribution over the remaining sequences was constructed by weighting sequences in proportion to their hamming distance from other sequences.

\begin{small}
\begin{sc}
\begin{tabular}{lrrrrr}
\toprule
& \multicolumn{4}{c}{Number of sequences in MSA} \\
 Sequence Sim  & 60 & 300 & 600 &  300 (ensem) &600 (ensem) \\
\midrule
 1.0 & 0.33 & 0.35 & 0.36 & 0.36 & 0.43 \\
 0.7 & 0.31 & 0.34 & 0.37 & 0.40 & 0.38\\
 0.4 & 0.33 & 0.38 & 0.39 & 0.33 & 0.38\\
 All & 0.32 & 0.37 & 0.37 & 0.37 & 0.42\\
\bottomrule
\end{tabular}
\end{sc}
\end{small}

\end{document}